# Cascaded optical fiber link using the Internet network for remote clocks comparison

**Nicola Chiodo,[1] Nicolas Quintin,[1] Fabio Stefani[2,1], Fabrice Wiotte,[1] Emilie Camisard,[3] Christian Chardonnet,[1] Giorgio Santarelli,[4] Anne Amy-Klein,[1,*] Paul-Eric Pottie,[2] Olivier Lopez[1]**

[1]*Laboratoire de Physique des Lasers, Université Paris 13, Sorbonne Paris Cité, CNRS, 99 Avenue Jean-Baptiste Clément, 93430 Villetaneuse, France*
[2]*Laboratoire National de Métrologie et d'Essais–Système de Références Temps-Espace, UMR 8630 Observatoire de Paris, CNRS, UPMC, 61 Avenue de l'Observatoire, 75014 Paris, France*
[3]*RENATER, 23-25 rue Daviel, 75013 Paris*
[4]*Laboratoire Photonique, Numérique et Nanosciences, UMR 5298 Université de Bordeaux , Institut d'Optique Graduate School and CNRS, 1, Rue F. Mitterand, 33400 Talence, France*
[*]*amy@univ-paris13.fr*

**Abstract:** We report a cascaded optical link of 1100 km for ultra-stable frequency distribution over an Internet fiber network. The link is composed of four spans for which the propagation noise is actively compensated. The robustness and the performance of the link are ensured by five fully automated optoelectronic stations, two of them at the link ends, and three deployed on the field and connecting the spans. This device coherently regenerates the optical signal with the heterodyne optical phase locking of a low-noise laser diode. Optical detection of the beat-note signals for the laser lock and the link noise compensation are obtained with stable and low-noise fibered optical interferometer. We show 3.5 days of continuous operation of the noise-compensated 4-span cascaded link leading to fractional frequency instability of 4x10-16 at 1-s measurement time and 1x10-19 at 2000 s. This cascaded link was extended to 1480-km with the same performance. This work is a significant step towards a sustainable wide area ultra-stable optical frequency distribution and comparison network at a very high level of performance.

OCIS codes: (120.3930) Metrological instrumentation; (060.2360) Fiber optics links and subsystems; (140.0140) Lasers and laser optics; (120.5050) Phase measurement.

## 1. Introduction

The transfer of ultra-stable frequency signals between distant locations is required by many applications, mostly for time-frequency metrology and high-sensitivity tests of fundamental physics. Remote comparison of the best optical clocks requires ultra-stable frequency transfer with a resolution of at least $10^{-18}$ after a few hours, which is far from being obtained with satellite links or transportable clocks [1]. By contrast, coherent optical fiber links gives the possibility to disseminate an ultra-stable frequency with a resolution below $10^{-19}$ on a few hundreds of km or more, thanks to an active compensation of the fiber [2-8]. A record distance of 1840 km was established in Germany with a relative frequency stability of $3x10^{-15}$ at 1 s averaging time and $10^{-19}$ after a few hundreds of s [9]. These experiments demonstrated that the transferred frequency shows no bias from the expected value with a relative uncertainty of $3x10^{-19}$. Record long-term stability at the $10^{-20}$ level was demonstrated on slightly shorter optical links [5, 10, 11].

In order to take advantage of the existing fiber network already connecting research facilities and universities via the National Research & Education Networks (NRENs), we have demonstrated that coherent optical links can be implemented on telecommunication fibers with data traffic, using dense wavelength division multiplexing (DWDM) technology [12]. However long distance Internet fibers use unidirectional optical equipment uncompliant with coherent fiber links, and hence this equipment has to be bypassed. Due to a higher number of connectors, they exhibit also larger losses and substantial parasitic reflections, as compared to dedicated fibers. We have thus developed a dedicated approach providing the optical regeneration of the signal, thanks to a repeater laser station (RLS), that our groups introduced in [13]. Optical regeneration is fully compatible with Internet network and allows us to even improve the rejection amplitude and bandwidth of the phase noise compensation, that are limited by the propagation delay through the fiber [2]. This way, we can build a long-haul cascaded link of successive spans over the Internet network. This approach also opens the way to the realization of a wide-area scalable metrological network.

In this article we will first show the optical fiber link we have implemented using the fibers of the French NREN RENATER coexisting with data traffic. We will also describe the operation principle of a cascaded link and show its practical implementation using four spans over 1100-km. In a second section, we present the architecture of a repeater station and its main functionalities. The experimental results are given in a third part, showing a very low instability of $4x10^{-16}$ at 1-s integration time. We finally present the results we obtained by extending the link to the German border in Strasbourg, totalizing a fiber length of 1480 km.

## 2. Cascaded optical links on Internet fiber network

An ultra-stable optical link is designed to reproduce with the best fidelity at the output end the frequency of an ultra-stable signal coupled at the input end of the fiber. However the optical phase is disturbed by the fluctuations of the optical path arising from the fiber temperature, mechanical and acoustical fluctuations. To overcome this limitation we measure the round-trip phase fluctuations and compensate them in real time.

In this paper we are first demonstrating an optical link of 1100 km between Villetaneuse (in the northern suburb of Paris) and Nancy (in the east of France) and back, using two parallel fiber spans, as sketched in Fig. 1. This link is part of a future link connecting Paris and Braunschweig, where the French and German National Metrology Institutes (NMIs) are located respectively [10]. This link is also part of the French project REFIMEVE+ to deploy a national fiber network between twenty research laboratories in order to disseminate an ultra-stable frequency signal for a wide range of applications [5, 14].

The deployment of such an optical link requires first the access to a pair of fibers which is difficult and expensive. Noticing that universities and research facilities are already interconnected by fiber networks for Internet data transmission, we proposed in 2009 to implement optical fiber links embedded in public telecommunication networks [12]. With this method, the ultra-stable optical signal shares the same fiber as the data traffic, using Dense Wavelength Division Multiplexing (DWDM). It is transmitted on a dedicated frequency channel of the DWDM grid leaving the rest of the optical spectrum available for data transmission. This technique using a science channel is compatible with active noise compensation providing slightly modifying the fiber network and setting-up specific equipment [12, 13]. Unlike optical-encoded data in telecommunication network, the ultra-stable optical signal needs to propagate from one end to the other end of the link in both directions in the same fiber in order to measure and compensate at best the phase fluctuations along the propagation. This implies that all the optical equipment, including online unidirectional Erbium Doped Fiber Amplifiers (EDFA) and DWDM platforms (used at frontend of network switches and routers) are bypassed. This is accomplished by installing optical add/drop multiplexers (OADM), able to inject or extract one single channel from the DWDM flux with very low insertion losses. In addition custom-made fully bidirectional amplifiers should be deployed in order to enable simultaneous forward and backward amplification.

The optical link Villetaneuse-Nancy-Villetaneuse was implemented using fiber spans of the National Research and Education Network of RENATER, each span consisting of two parallel fibers (the up and down fibers of the telecommunication network). The metrological signal is using the frequency channel centered at 1542.14 nm (44$^{th}$ channel of the International Telecommunication Union DWDM grid). Except for two 35-km dedicated fiber spans, the link is used simultaneously for data transfer. For the major part of the link, between Paris and Nancy and back (505-km of the total 550-km) the Internet data are transmitted on the neighboring frequency channel just 100 GHz away. Fig 2 displays a scheme of the link span from Reims to Nancy, which includes two network nodes and three in-field amplification sites. Eight OADMs are used to bypass the telecommunication equipment, and three bidirectional EDFAs are set-up in the huts. Detailed configurations of the link between Villetaneuse and Reims can be found in reference [4]. In total the link to Nancy and back to Villetaneuse is equipped with 32 OADMs and 12 EDFAs.

The total attenuation of the 1100-km fibers to Nancy and back is close to 320 dB. The average losses are about 0.29 dB/km, which is significantly greater than the 0.2 dB/km attenuation of the fiber at 1550 nm. This is mainly due to the high numbers of connectors, which induce also straight reflections. Since we have no access to the fibers, we are unable to reduce these reflections and losses, in contrast to optical links using dedicated fiber. Due to these reflections, we have observed that the optical gains of the bi-directional EDFAs must be kept around 15 dB to avoid self-oscillation. Thus the link attenuation cannot be fully compensated by far using only EDFAs. For instance assuming EDFAs had been installed at each node and huts, one would obtain a net optical budget below -100 dB and the active noise compensation cannot work. Brillouin amplifiers, that can achieve high gain up to 60 dB in a narrow band typically below 10 MHz, can be used to better amplify the signal [10, 15], however the optical pump power in excess of 10 mW required for achieving high gain is hardly compatible with a public optical telecommunication network operation.

In order to overcome this limitation, we have implemented a cascaded link: the link is divided in a few spans connected with repeater laser stations (RLS), as demonstrated in [4, 13]. RLSs are able to repeat the optical phase from one span to the next span with optimal power. The main element of a station is a narrow linewidth laser diode which is phase-locked to the incoming signal. This laser is injected in the next span and sent back to the previous station in order to detect and compensate the noise of the previous span. The round-trip signal of the following span is also detected in order to correct the propagation noise to the next station.

Here we implement a cascaded link of four spans, as depicted on Fig. 1. The first span connects Villetaneuse to Reims (270 km), the second one Reims to Nancy (280 km) and the third and fourth spans loop back from Nancy to Reims and then Reims to Villetaneuse. It includes two stations at the link input and output ends in Villetaneuse, two stations in Reims and one in Nancy. All the repeater stations are operated remotely.

This scheme, beyond enabling a high equivalent optical gain, low noise and narrow band amplification, enhance the noise compensation. In fact, both the noise rejection factor and the control bandwidth decrease with the propagation delay [2]. When dividing the link into successive spans, the propagation delay in each fiber span is reduced and thus the noise compensation is improved. Assuming that the free running fiber noise power scales as the length of each subsection, and the noise distribution is homogeneous along the link, one expects that the relative frequency stability of a cascaded link, divided into N segments, is improved by a factor equal to $\sqrt{N}$ and that the correction bandwidth is improved by a factor N [16, 17]. This may improve essentially short-term and mid-term stability (below a few thousands s), since the long-term stability is usually limited by the detection noise floor. A larger correction bandwidth is critical when the transferred signal may be used for remote laser stabilization for instance [18]. Moreover the use of stations to repeat the signal gives the possibility to extract the ultra-stable signal for a local user and thus to distribute it to several users simultaneously using a single fiber link. This is thus a key element for implementing a wide-scale metrological fiber network.

## 3. The repeater laser station.

The heart of a cascaded link is the repeater laser station, which is depicted in Fig 3. The principle of operation of a repeater laser station has been reported in [4]. We are reminding here only the main features, and focusing on the updated features. The core concept is the phase-lock of a narrow-line (<5kHz) laser diode set up in the remote site N to the upcoming signal of the link N (which connects site N-1 to site N). For that purpose, the beat-note between the laser diode and the link N signal is detected through a low-noise Michelson-type interferometer. The incoming signal power can be as low as a few nW, with a signal-to-noise ratio of about 90 dB in 1 Hz bandwidth. The phase fluctuations of the link are then copied onto the optical phase of the station laser. The signal of this optical local oscillator is then sent back to the previous station N-1, and compared to the signal of the station N-1 using a strongly unbalanced Michelson-type interferometer. This way, the round trip propagation noise of the link N is detected and corrected using an AOM. Thus the laser diode optical phase of the remote site N is copying the laser optical phase from the previous station N-1. The next span (N+1) is seeded with this regenerated laser light. The noise of the next span is compensated using another Michelson-type interferometer in order to compare the signal which retraces back from station N+1 with the local laser diode signal of station N. Corrections are applied using an AOM. Last but not least, the remote laser can be used to feed an auxiliary output, which delivers the ultra-stable signal to a local user using an auxiliary short stabilized optical link.

### *3.a Passive compensation of the local RF oscillator fluctuations*

The inaccuracy and instability of the local RF oscillator are issues in such a cascaded link. It can limit the stability and the accuracy of the transferred signal. Each phase-lock loop or AOM is indeed shifting the laser frequency with a radio-frequency synthesized from a local RF oscillator. Thus the latter instabilities are added to the ultra-stable signal. We circumvent this problem by choosing the AOMs and phase-lock loops frequencies such that the instabilities of the local oscillator cancel out. The laser phase-lock loop is driven with a frequency $f_N$ and the link phase-lock loop with twice this frequency. A third AOM ($AOM_1^N$ in Fig. 3) driven at $f_N/2$ was inserted in the station. It compensates the frequency shift induced by the phase-lock loops and the AOM ($AOM_2^N$ on Fig. 3) used for the link N+1 noise compensation. Thus the signal at the station N output is independent of the local RF oscillator of station N.

The same compensation technique is used for the local user output of station N. In that case, the compensation loop of the user link should simply be driven with frequency $-f_N$.

### *3.b Realization of the repeater laser station*

The signal processing for such a long haul optical link is quite specific. For each span, the propagation noise can be very large, leading to phase excursions of the upcoming signal up to 100 radians, with occasional peaks as high as a few thousands radians. These are imprinted onto the laser by a phase-lock loop, where the laser acts as an optical tracking oscillator. The phase-lock loop is offset by a fixed frequency, so that we can detect specifically the heterodyne beat-note of the link N round-trip signal and reject parasitic beat notes arising by stray reflections. The same challenge occurs for stabilizing the link N+1. For that purpose, a RF oscillator is phase-locked to the round-trip signal. Such a so-called tracking oscillator is acting as a high-gain narrow-bandwidth amplifier and limiter. Then the tracked signal is digitally divided using a Direct Digital Synthesizer. The phase of the tracked and divided signal is then compared to that of a low-frequency RF oscillator signal, to generate the error signal. The latter is input at the servo-loop that controls the frequency of the AOM used to apply the noise correction. Using a well-chosen division factor of 150 to 1500, the phase excursion is maintained in the range $-2\pi$ to $+2\pi$. All these phase-lock loops are automatically operated by a micro-controller.

In order to optimize the beat-note power between the upcoming signal of the link N and the repeater laser, the polarization states must be matched. As the state of polarization out of the link is slowly varying with time, we insert a fiber polarization controller driven by a micro-controller that measures and optimizes the beat-note signal strength. Thanks to the Faraday mirrors used in the interferometers at the two ends of the link N, the polarization states are therefore also matched for the roundtrip beat-note signal.

One of the key elements of the repeater station is the interferometric setup, including the two interferometers used to detect the round-trip beat-note and the beat-note between the laser diode and the link signal. It is realized with spliced fiber-optic components and housed in a compact optical module including one AOM. A careful design ensures that the optical phase difference between the input and output signals is as insensitive as possible to thermal fluctuations and gradient in the station. A detailed analysis and an experimental benchmarking are reported in [19]. We successfully realize interferometric setups with temperature sensitivity as low as +1fs/K which corresponds to a 7-fold improvement compared to our previous setup.

The whole repeater station is steered by an embedded computer, which communicates with the microcontrollers and is accessible through Internet protocol. It gives us the possibility to operate remotely the laser station and optimize the ultra-stable signal transfer. We are also remotely recording the error signal of the compensation loop of the link stabilization, Calculating the Fast Fourier Transform (FFT) of this link error signal, we can optimize the gain and thus the bandwidth of the link correction loop and detect eventual unwanted oscillations. In addition, we are able to detect and time tag the cycle slips of the three phase-lock loops. The events corresponding to an optimization of the polarization states are also tagged.

## 4. Results and discussions

We first use these stations on the 1100-km cascaded link to Nancy (Fig. 1). The optical signal injected in the link is provided by a laser stabilized to an ultra-stable cavity, which is located at laboratory SYRTE and transferred to LPL through a dedicated optical fiber of 43 km [16]. In order to characterize the end-to-end stability of this 1100-km link, we measure the beat-note signal between the link's input and output ends with a short link connecting the two laser stations at LPL. The beat-note signal is filtered with a band pass filter of bandwidth below 2 MHz, amplified, and then counted with a dead-time free frequency counter [K+K Messtechnik FXE80]. We operate the counter in Lambda-mode with a gate time of 1 s [20, 21]. The beat-note signal is also down-converted to baseband for a Fast Fourier analysis.

We first present on Fig. 4a the in-loop optical phase of the round-trip compensated signal of each spans for the 1100 km link. We can observe that the residual phase noise increases by each span. This phenomenon is also seen on Fig. 4b, which displays the phase noise power spectral density of each span. At each span, the fiber noise fraction that is above the correction bandwidth is left uncorrected. This remaining noise is copied onto the next span, and cannot be corrected for. As a matter of fact the last span has to deal with the cumulated uncorrected phase noise of the three previous spans plus the noise of the last span link. The phase noise excursion is so large that we need to divide the tracked signal of the spans 3 and 4 (see Fig. 1) by a factor 1500. In these conditions, signal-to-noise ratio is a critical parameter, as the number of cycle slips depends exponentially on the signal-to-noise ratio of the signal that enters the phase comparator. In such a long haul cascaded link, we had to finely adjust the gain parameters of the four tracking oscillators and four correction loops, the input gains, and the optical gain of the sixteen EDFA amplifiers.

When all the gain parameters are optimized, we are able to operate the cascaded link continuously for several days. We present on Fig. 5 the end-to-end optical phase fluctuations for a 3.5 days measurement (red curve). Only 73 data points coinciding with cycle slips were removed, most of them corresponding to an adjustment of the polarization. The phase variations remain in a 20 rad range along the measurement and its fluctuations are less than 2 rad in the short term. We did not study these long-term oscillations and drift, which may be attributed to temperature change in the station interferometers, long-term autocorrelation noise of the ultra-stable laser and eventually effects of Polarisation Mode Dispersion (PMD). The corresponding power spectral density, measured by a FFT analyzer at the local end, is reported in Fig. 4b (red curve). It is below -10 $dBrad^2/Hz$ between 1 and 10 Hz. The servo peak around 100 Hz is compatible with the bandwidth limit of 180 Hz given by the propagation delay of 1.4 ms in the longer span of 280 km. It is a significant reduction of the free-running fiber noise, which is above 20 $dBrad^2/Hz$ at 1 Hz for the 540-km one-span link from Villetaneuse to Reims and back [13]. Note that the free-running fiber noise of each span cannot be simply evaluated from the correction signal, because of the instabilities of the RF oscillator of each in-field stations.

We finally present on Fig. 6 the instability of the relative frequency fluctuations of the end-to-end phase measurements, calculated using the modified Allan deviation (red filled squares) [21]. With this 1100 km LPL-Nancy-LPL link, we demonstrate a frequency stability of $4\times10^{-16}$ at 1 s integration time, which averages down to $5\times10^{-20}$ at 60 000 s integration time. This stability performance allows us the comparison of the best optical clocks after only 100 s [22, 23]. For shorter data samples of around 12 hours, the long-term stability reaches a few $10^{-20}$ after 10 000 s.

Finally, the accuracy of the frequency transfer was evaluated by calculating the mean value of the end-to-end beat note frequency offset. Following [5, 10], we estimate its statistical fractional uncertainty as the long-term overlapping Allan deviation at 64000 s of the data set. We obtained for the LPL-Nancy-LPL link a mean frequency offset of $+2.9\times10^{-20}$ with a statistical uncertainty of $1.3\times10^{-19}$.

## 5. Extension to a 1480-km link to Strasbourg

In a second step, we extended the cascaded link to Strasbourg by moving the repeater station from Nancy to Strasbourg. We also added EDFAs in Nancy and in a hut between Nancy and Strasbourg to partially compensate the 50 dB additional losses. Besides these four EDFAs, this extension adds up eight OADMs.

Due to the additional fiber sections and increased number of EDFAs, long-run stable operation of this link was more difficult than with the Nancy link. Unfortunately, we found out that EDFAs gains are not that stable and can vary up to 6 dB, according to the room temperature variations. This problem is not that trivial as the input optical signal can be very weak, and link stabilization is more difficult to implement. At present this restricts long-term operation of the optical link, requiring unavoidable readjustment sessions.

Fig. 5 displays the end-to-end optical phase fluctuations of the 1480 km link (blue curve). For this best-case measurement, recorded on a period of 12 hours, we removed only 4 cycle slips. Phase fluctuations are slightly degraded compared to the shorter 1100-km link. For longer operation time, we observe that the link signal arriving at the third station exhibits non-stationary fluctuations which degrade the operation of this station. We attribute this degradation partly to the additional EDFAs and the difficulty to adjust their gain at long-term, as mentioned above.

We present on Fig. 6 the stability of the relative frequency fluctuations of the 1480-km link (blue circles), which is comparable to that of the Nancy link. The modified Allan deviation is $6\times10^{-16}$ at 1 s integration time and $2\times10^{-20}$ at 8 000 s of integration time for this best case run. For typical measurements, the long-term stability is $10^{-19}$ or below. Concerning the accuracy, we obtained also similar figures than with the Nancy link, namely a mean frequency offset of $-4.8\times10^{-20}$ with a statistical uncertainty of $9\times10^{-20}$ given by the long-term overlapping Allan deviation at 8000 s.

For both the Nancy and the Strasbourg links, we observe that the fiber noise is not stationary and can increase significantly, probably due to human activities on the link, as work road for instance. This excess of noise increases the cycle slip rate and degrades the stability. The stability can also be improved for instance after a maintenance of the fiber network.

## 6. Conclusion

To the first time, we report a 4-span cascaded link of 1100 km, with record relative frequency stability of $4\times10^{-16}$ at 1 s integration time and $5\times10^{-20}$ at 60 000 s integration time. We show that this cascaded link is operated with very low cycle slip rate over days of continuous stable operation. This performance can be achieved only after a careful adjustment of all the gains of the amplifiers and phase-lock loops. We point out that the extension to 1480 km degrades the robustness of the cascaded compensated link. However, both links gives us the possibility to compare future optical clocks at the $10^{-19}$ level after only a few hours.

These achievements demonstrate the strength of the cascaded approach, which allows us to overcome both the classical limit of noise rejection by the optical delay and the attenuation of the link. The demonstrated technique is operated over active telecommunication network with parallel data traffic. Our results show that ultra-high stability and accuracy can be achieved on such a network backbone. By comparison with the achievement on dedicated fiber, we show that the parallel transmission of data did not impact the performance of stability and accuracy of the link, at least to the level of $5\times10^{-20}$. It opens the way to an extension of this dissemination technique over national and international academic and research networks, for a wide range of applications.

This work is a major step of the French project named REFIMEVE+ [14]. REFIMEVE+ is developing a wide national infrastructure where a reference optical signal generated at SYRTE is to be distributed to about 20 academic and institutional users using the RENATER network. The link from Paris to Strasbourg realizes actually the first branch of the REFIMEVE network with laboratory prototypes. In addition, this project targets transnational links towards Germany, UK and Italy. The link end in Strasbourg opens the way to optical clock comparisons between SYRTE and PTB in a near future.

## Acknowledgments

This work would not have been possible without the support of the GIP RENATER. The authors are deeply grateful to L. Gydé, T. Bono from GIP RENATER. We also acknowledge A. Kaladjian from LPL for technical support. We acknowledge funding support from the Agence Nationale de la Recherche the Agence Nationale de la Recherche (ANR blanc LIOM 2011-BS04-009-01, Labex First-TF ANR 10 LABX 48 01, Equipex REFIMEVE+ ANR-11-EQPX-0039), the European Metrology Research Programm (contract SIB-02 NEAT-FT), IFRAF-Conseil Régional Ile-de-France, CNRS with Action Spécifique Gravitation, Références, Astronomie, Métrologie (GRAM).

# References and links

1. A. Bauch, J. Achkar, S. Bize, D. Calonico, R. Dach, R. Hlava, L. Lorini, T. Parker, G. Petit, D. Piester, K. Szymaniec, and P. Uhrich, "Comparison between frequency standards in Europe and the USA at the $10^{-15}$ uncertainty level," Metrologia 43, 109 (2006).
2. N. R. Newbury, P. A. Williams, and W. C. Swann, "Coherent transfer of an optical carrier over 251 km," Optics Letters 32, 3056 (2007).
3. K. Predehl, G. Grosche, S. M. F. Raupach, S. Droste, O. Terra, J. Alnis, T. Legero, T. W. Hänsch, T. Udem, R. Holzwarth, and H. Schnatz, "A 920-kilometer optical fiber link for frequency metrology at the $19^{th}$ decimal place," Science 336, 441-444 (2012).
4. O. Lopez, A. Haboucha, B. Chanteau, C. Chardonnet, A. Amy-Klein, and G. Santarelli, "Ultra-stable long distance optical frequency distribution using the Internet fiber network," Optics Express 20, 23518-23526 (2012).
5. O. Lopez, F. Kéfélian, H. Jiang, A. Haboucha, A. Bercy, F. Stefani, B. Chanteau, A. Kanj, D. Rovera, J. Achkar, C. Chardonnet, P. E. Pottie, A. Amy-Klein, and G. Santarelli "Frequency and time transfer for metrology and beyond using telecommunication network fibres," Comptes Rendus Physique 16, 531-539 (2015).
6. D. Calonico, E. K. Bertacco, C. E. Calosso, C. Clivati, G. A. Costanzo, M. Frittelli, A. Godone, A. Mura, N. Poli, D. V. Sutyrin, G. Tino, M. E. Zucco, and F. Levi, "High-accuracy coherent optical frequency transfer over a doubled 642-km fiber link," Applied Physics B 117, 979-986 (2014).
7. L. Sliwczynski, P. Krehlik, L. Buczek, and M. Lipinski, "Frequency transfer in electronically stabilized fiber optic link exploiting bidirectional optical amplifiers," IEEE Trans. Instr. Meas. 61, 2573-2580 (2012).
8. F. L. Hong, M. Musha, M. Takamoto, H. Inaba, S. Yanagimachi, A. Takamizawa, K. Watabe, T. Ikegami, M. Imae, Y. Fujii, M. Amemiya, K. Nakagawa, K. Ueda, and H. Katori, "Measuring the frequency of a Sr optical lattice clock using a 120 km coherent optical transfer," Optics Letters 34, 692 (2009).
9. S. Droste, F. Ozimek, T. Udem, K. Predehl, T. W. Hänsch, H. Schnatz, G. Grosche, and R. Holzwarth, "Optical-Frequency Transfer over a Single-Span 1840 km Fiber Link," Physical Review Letters 111, 110801 (2013).
10. S. M. F. Raupach, A. Koczwara, and G. Grosche, "Brillouin amplification supports 1 x $10^{-20}$ accuracy in optical frequency transfer over 1400 km of underground fibre," Physical Review A 92, 021801(R) (2015).
11. O. Lopez, N. Chiodo, F. Stefani, F. Wiotte, N. Quintin, A. Bercy, C. Chardonnet, G. Santarelli, P.-E. Pottie, and A. Amy-Klein, "Cascaded optical link on a telecommunication fiber network for ultra-stable frequency dissemination," Proc. SPIE. 9378, Slow Light, Fast Light, and Opto-Atomic Precision Metrology VIII, (2015).
12. F. Kéfélian, O. Lopez, H. Jiang, C. Chardonnet, A. Amy-Klein, and G. Santarelli, "High-resolution optical frequency dissemination on a telecommunications network with data traffic," Optics Letters 34, 1573 (2009).
13. O. Lopez, A. Haboucha, F. Kéfélian, H. Jiang, B. Chanteau, V. Roncin, C. Chardonnet, A. Amy-Klein, and G. Santarelli, "Cascaded multiplexed optical link on a telecommunication network for frequency dissemination," Optics Express 18, 16849 (2010).
14. Refimeve+, "project www.refimeve.fr/."

15. O. Terra, G. Grosche, and H. Schnatz, "Brillouin amplification in phase coherent transfer of optical frequencies over 480 km fiber," Optics Express 18, 16102 (2010).
16. H. Jiang, F. Kéfélian, S. Crane, O. Lopez, M. Lours, J. Millo, D. Holleville, P. Lemonde, C. Chardonnet, A. Amy-Klein, and G. Santarelli, "Long-distance frequency transfer over an urban fiber link using optical phase stabilization," J. Opt. Soc. Am. B 25, 2029-2035 (2008).
17. M. Fujieda, M. Kumagai, and S. Nagano, "Coherent microwave transfer over a 204-km telecom fiber link by a cascaded system," IEEE Transactions on Ultrasonics, Ferroelectrics, and Frequency Control 57, 168-174 (2010).
18. B. Argence, B. Chanteau, O. Lopez, D. Nicolodi, M. Abgrall, C. Chardonnet, C. Daussy, B. Darquié, Y. Le Coq, and A. Amy-Klein, "Quantum cascade laser frequency stabilization at the sub-Hz level," Nat Photon 9, 456-460 (2015).
19. F. Stefani, O. Lopez, A. Bercy, W.-K. Lee, C. Chardonnet, G. Santarelli, P.-E. Pottie, and A. Amy-Klein, "Tackling the limits of optical fiber links," Journal of the Optical Society of America B 32, 787-797 (2015).
20. E. Rubiola, "On the measurement of frequency and of its sample variance with high-resolution counters," Review of Scientific Instruments 76, 054703-054706 (2005).
21. S. T. Dawkins, J. J. McFerran, and A. N. Luiten, "Considerations on the measurement of the stability of oscillators with frequency counters," Ultrasonics, Ferroelectrics, and Frequency Control, IEEE Transactions on 54, 918-925 (2007).
22. T. L. Nicholson, S. L. Campbell, R. B. Hutson, G. E. Marti, B. J. Bloom, R. L. McNally, W. Zhang, M. D. Barrett, M. S. Safronova, G. F. Strouse, W. L. Tew, and J. Ye, "Systematic evaluation of an atomic clock at 2 x $10^{-18}$ total uncertainty," Nat Commun 6, 6896 (2015).
23. I. Ushijima, M. Takamoto, M. Das, T. Ohkubo, and H. Katori, "Cryogenic optical lattice clocks," Nat Photon 9, 185-189 (2015).

# Figures

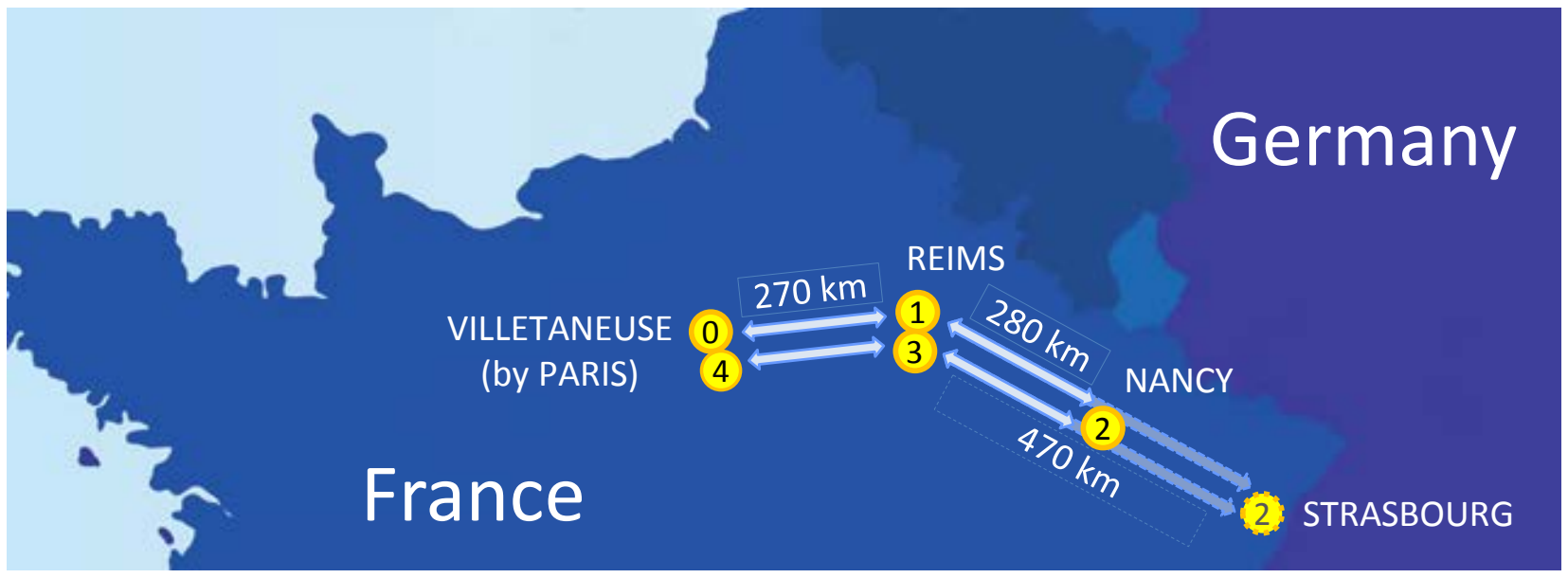

Fig. 1. Scheme of the four-spans cascaded optical link using the RENATER fiber network between Villetaneuse and Nancy, and its extension to Strasbourg.

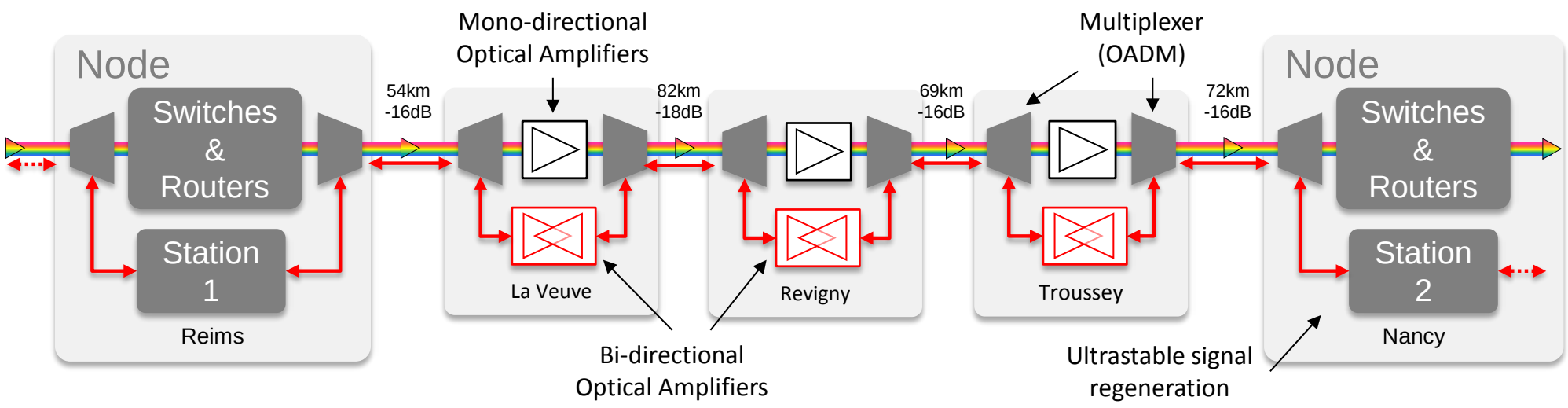

Fig. 2. Detailed scheme of the second span between Reims and Nancy.

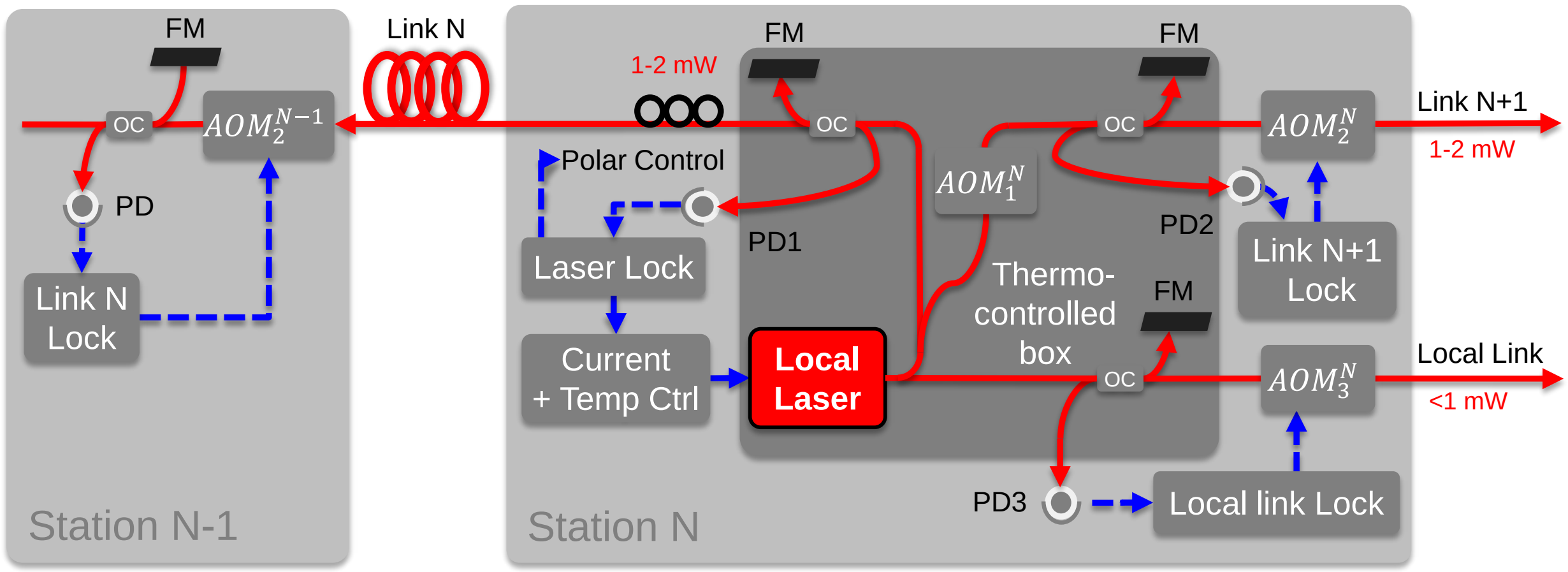

Fig. 3. Schematic of the Nth repeater station, OC: optical coupler, FM: Faraday mirror, PD: photodiode, PC: polarization controller, AOM: acousto-optic modulator.

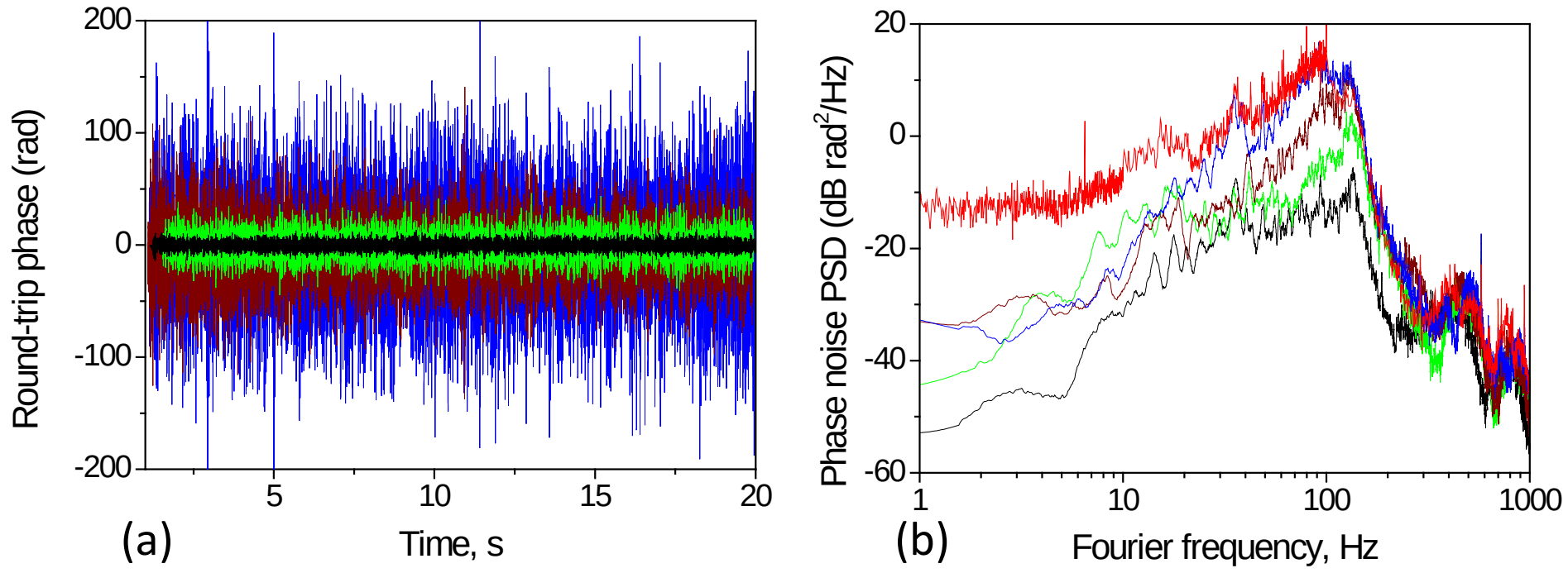

Fig. 4. a) Phase fluctuations along time of each span of the compensated cascaded link (black, green, brown and blue curves for the spans 1 to 4, as depicted on Fig. 1 calculated from the in-loop error signal, b) Phase noise spectral density of the successive spans of the cascaded link (same colors as for (a)) calculated from the in-loop error signal (and divided by two) and the whole cascaded link (red curve) calculated from the end-to-end phase signal.

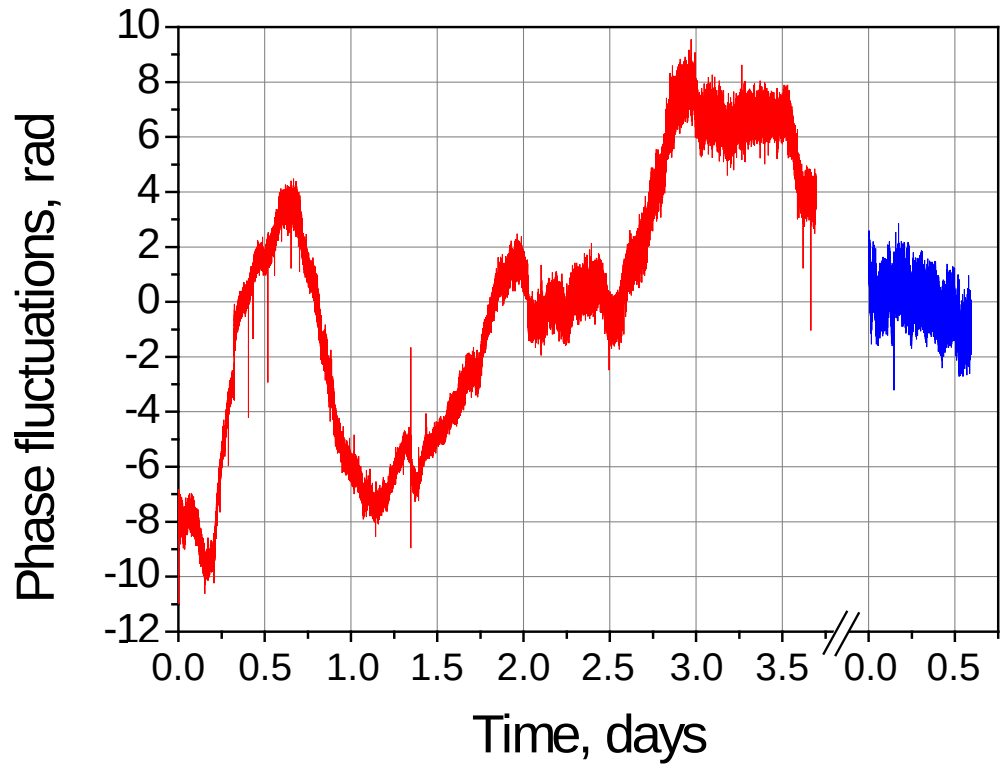

Fig. 5. End-to-end phase fluctuations along time of the Villetaneuse-Nancy-Villetaneuse 1100-km compensated optical link (red curve) and its extension to Strasbourg (blue curve).

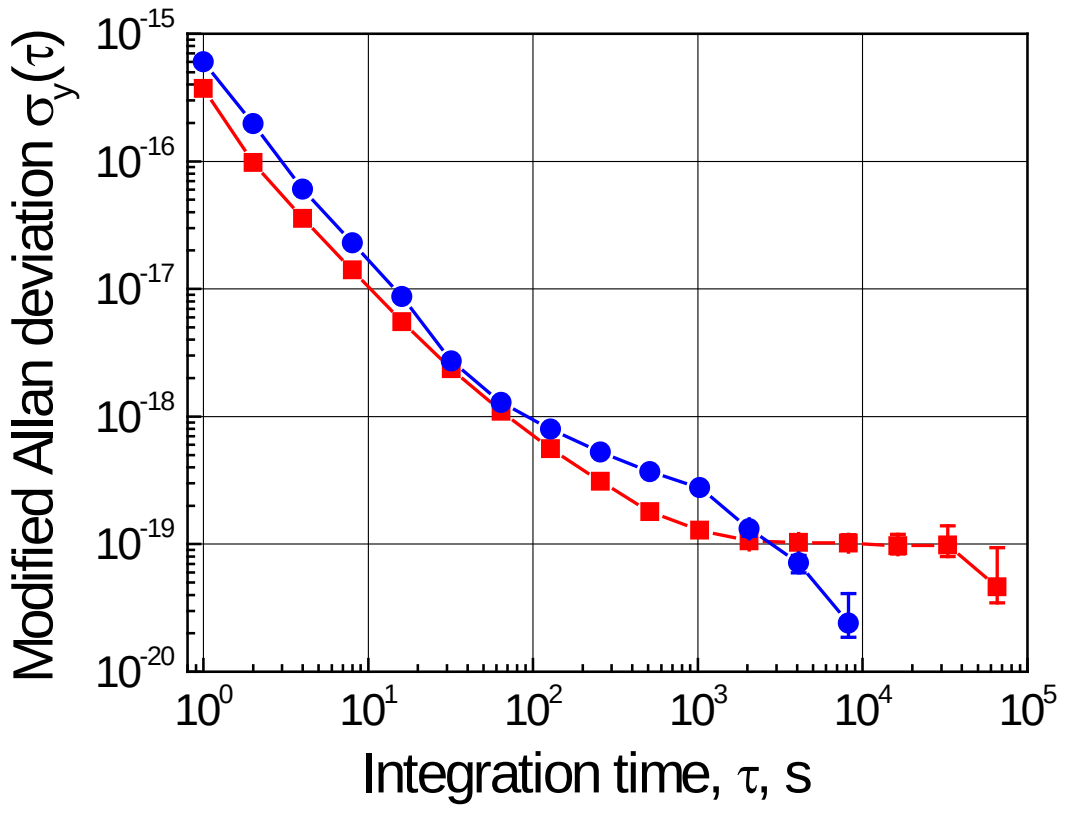

Fig. 6. Fractional frequency instability versus averaging time of the Villetaneuse-Nancy-Villetaneuse 1100-km compensated optical link (red squares) and of its extension to Strasbourg (blue circles). The stabilities are calculated from Λ-type data using modified Allan deviation.